\documentclass[a4paper,11pt]{article}
\usepackage{pos}

\newcommand{\as}{\alpha_{\mathrm{S}}}

\newcommand{\cic}{\sum_{\mathrm{cyc.}{123}}}       

\newcommand{\dug}{\equiv}
\newcommand\eqcs{\stackrel{\mathsf{cs}}{=}}
\newcommand{\eik}{\mathrm{eik}}             
\newcommand{\g}{\mathrm{g}}              
\newcommand{\half}{{\textstyle\frac{1}{2}}}
\newcommand{\imp}{\Longrightarrow}
\newcommand{\J}{\boldsymbol{J}}             
\newcommand{\ket}[1]{| #1 \rangle}

\newcommand{\M}{{\cal M}}               
\newcommand{\ord}[1]{\mathcal{O}\left(#1\right)}
\newcommand{\pol}{\varepsilon}
\newcommand{\qb}{\bar{\mathrm{q}}}       

\renewcommand{\S}{\mathcal{S}}             
\newcommand{\stro}{\mathsf{seo}}  
\newcommand{\T}{\boldsymbol{T}}                   
\newcommand{\tr}{\mathrm{tr}}            
\newcommand{\ugd}{=\hspace{-3.2pt}\raisebox{0.37pt}{:}\hspace{3pt}} 
\newcommand{\ui}{\mathrm{i}}             
\newcommand{\ul}[1]{\underline{#1}}

\renewcommand{\qb}{\bar{q}}
\renewcommand{\stro}{\mathsf{SEO}}  

\title{Triple (and quadruple) soft-gluon radiation in QCD hard scattering}

\author*[a,b]{Dimitri Colferai}
\author[b]{Stefano Catani}
\author[a]{Alessandro Torrini}

\affiliation[a]{Dipartimento di Fisica e Astronomia, Universit\`a di Firenze,\\
  Via Sansone 1, Sesto Fiorentino, Italy}

\affiliation[b]{INFN, Sezione di Firenze,\\
  Via Sansone 1, Sesto Fiorentino, Italy}

\emailAdd{dimitri.colferai@unifi.it}
\emailAdd{catani@fi.infn.it}
\emailAdd{torrini@fi.infn.it}

\abstract{We consider the radiation of three soft gluons in a generic
  process for multiparton hard scattering in QCD.  In the soft limit
  the corresponding scattering amplitude has a singular behaviour that
  is factorized and controlled by a colorful soft current.  We compute
  the tree-level current for triple soft-gluon emission from both
  massless and massive hard partons.  The three-gluon current is
  expressed in terms of maximally non-abelian irreducible
  correlations.  We compute the soft behaviour of squared amplitudes
  and the colour correlations produced by the squared current.  The
  radiation of one and two soft gluons leads to colour dipole
  correlations.  Triple soft-gluon radiation produces in addition
  colour quadrupole correlations between the hard partons.  We examine
  the soft and collinear singularities of the squared current in
  various energy ordered and angular ordered regions.  We discuss some
  features of soft radiation to all-loop orders for processes with two
  and three hard partons.  Considering triple soft-gluon radiation
  from three hard partons, colour quadrupole interactions break the
  Casimir scaling symmetry between quarks and gluons.  We also present
  some results on the radiation of four soft gluons from two hard
  partons, and we discuss the colour monster contribution and its
  relation with the violation (and generalization) of Casimir scaling.
  We also compute the first correction of $(1/N_c^2)$ to the eikonal
  formula for multiple soft-gluon radiation with strong energy
  ordering from two hard gluons.}

\FullConference{%
  Loops and Legs in Quantum Field Theory - LL2022,\\
  25-30 April, 2022\\
  Ettal, Germany
}

\begin{document}
\maketitle

\section{Introduction}

The subject of this talk is definitely on the legs' side rather than on the
loops' side.  It deals with tree-level emission of three and also four soft
gluons from any QCD hard scattering.

There are many motivations for studying soft gluon emissions from arbitrary
amplitudes.  First of all, high-precision LHC data require high precision in
theoretical predictions, both to test our present understanding of the Standard
Model and to discover (probably tiny) signals of new physics.  Furthermore, the
explicit knowledge of soft/collinear factorization of scattering amplitudes is
necessary in resummed calculations at next-to-next-to-next-to-leading (N$^3$L)
order.  In addition, calculation of large logarithmic terms can be used to
obtain approximated fixed-order results. Finally, soft/collinear factorization
provides the theoretical basis of parton shower algorithms for Monte Carlo event
generators.

\begin{figure}[hb]
  {\centering  \includegraphics[width=0.6\linewidth]{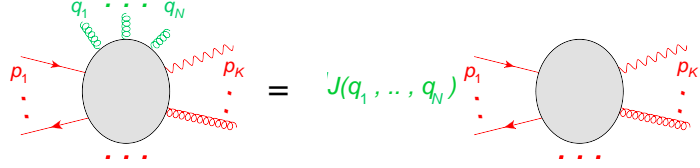}\\}
  \caption{Schematics of the soft factorization formula.\label{f:fact}}
\end{figure}

Our aim is to investigate the behaviour of a scattering amplitude when some
external gluons become soft.
We denote with $p$ the momenta of hard particles, and with $q$ those of the soft
gluons. When gluons become soft, the amplitude diverges, and the leading divergence
can be factorized as a singular soft current operator $J$
acting on the hard amplitude with the soft gluons removed, as depicted in
fig.~\ref{f:fact}.

\section{Soft currents\label{s:sc}}

Let me first consider the case of one photon emission in QED.  The soft current
is just a number which depends on the hard and soft momenta and on the charges
of the hard particles (eq.~\eqref{qed}).  A similar expression holds in
QCD~\cite{Bassetto:1984ik}, but the abelian charge is replaced by the non-abelian colour
matrices whose explicit expressions are in eq.~\eqref{qcd}
\begin{align}\label{qed}
  \text{QED:}\qquad J(q) &= J^\mu(q) \pol_\mu(q)
  = \sum_{k=1}^K \frac{e_k p_k^\mu}{p_k\cdot q} \pol_\mu(q)  \\
  \text{QCD:}\qquad J^a(q) &= J^{a,\mu}(q) \pol_\mu(q) = \sum_{k=1}^K
  \frac{g T^a_k p_k^\mu}{p_k\cdot q} \pol_\mu(q)\qquad
  \begin{cases}
    _{(T_q^a)_{bc} = t^a_{bc}}\\
    _{(T_{\bar{q}}^a)_{bc} = -t^a_{cb}}\\
    _{(T_g^a)_{bc} = \ui f^{abc}}
  \end{cases}
  \label{qcd} \\ \label{grav}
\text{GRAV:}\qquad J(q) &= J^{\mu\nu}(q) \pol_{\mu\nu}(q) = \sum_{k=1}^K
    \frac{\kappa p_k^\nu p_k^\mu}{p_k\cdot q} \pol_{\mu\nu}(q)
\end{align}
For physical amplitudes, with overall zero charge (colour singlets),
the currents are conserved:%
\footnote{The notation $\eqcs$ means equality of operators when acting on colour
singlet states.}
\begin{align}
  \sum_{k=1}^K e_k &= 0 \;, \qquad \sum_{k=1}^K T^a_k \eqcs 0
  \quad\imp\quad q_\mu J^{a,\mu} \eqcs 0
\end{align}
It is interesting to show that a soft current with the same structure exists in
gravity as well, see eq.~\eqref{grav}, with the colour charge replaced by the
4-momenta of the hard particles.

Also the current for two soft gluon emission is known since long~\cite{CaGr99}.
It can be conveniently expressed as the symmetric product of two 1-gluon
currents, plus a remainder $\Gamma(1,2)$ which represents the maximally
non-abelian correlation between the two soft gluons:
\begin{align*}
  J^{a_1 a_2}_{\mu_1 \mu_2}(q_1,q_2) &= J^{a_1}_{\mu_1}(q_1) \ast
  J^{a_2}_{\mu_2}(q_2) + \Gamma^{a_1 a_2}_{\mu_1 \mu_2}(q_1,q_2)
  \qquad\quad \left(A\ast B \dug \half(AB+BA)\right) \\
  \Gamma^{a_1 a_2}_{\mu_1 \mu_2}(q_1,q_2) &= \ui f^{a_1 a_2\, b} \sum_{k\in\text{hard}} T_k^b 
  \;\gamma_k^{\mu_1 \mu_2}(q_1,q_2) \\
  \gamma_k^{\mu_1 \mu_2}(q_1,q_2) &=
      \frac1{p_k\cdot(q_1+q_2)} \left\{
      \frac{p_k^{\mu_1}  p_k^{\mu_2}}{2\;p_k\cdot q_1}
      + \frac1{q_1\cdot q_2}
      \left(p_k^{\mu_1} q_1^{\mu_2}+\frac12 g^{\mu_1\mu_2}p_k\cdot q_2\right) \right\}
      - \bigl( 1 \leftrightarrow 2 \bigr) \;.
\end{align*}
The colour structure of $\Gamma$ is just a colour matrix contracted with a
structure constant.  This current is conserved, if applied on colour-singlet
states, even without contracting the remaining Lorentz index with the
polarization vector of the second gluon.  In the abelian case only the
independent product of single gluon currents survives.  In QCD however we have
colour correlation also from the symmetric product.  By the way, the 1-gluon
currents do not commute, since they contain colour matrices.

We have evaluated~\cite{CCT} the current for the emission of three gluons computing
the relevant Feynman diagrams with gluon insertions on the external lines, where
we use eikonal vertices on hard lines and exact vertices and propagators
elsewhere.  We find that the current, when acting on colour singlets, does not
depend on the gauge and is conserved.  It can conveniently be expressed as a
symmetric product of three 1-gluon currents, the symmetric product of one
1-gluon current and the 2-gluon non-abelian term, plus an irreducible, maximally
non-abelian remainder $\Gamma(1,2,3)$:
\begin{align}
  J(1,2,3) &= J(1) \ast J(2) \ast J(3)
  + \Big[\cic J(1)\ast\Gamma(2,3)\Big] + \Gamma(1,2,3) \\
  \Gamma(1,2,3) &= \sum_{k\in\text{hard}} T_k^b \cic \sum_s
  f^{a_1 a_2\, s}f^{s\, a_3 b} \;\gamma_k^{\mu_1 \mu_2 \mu_3}(q_1,q_2;q_3)
\end{align}
We can present the soft currents also in terms of the decomposition in
colour-ordered subamplitudes
  \begin{align*}
    M(1,\cdots,n) &= \sum_{\mathrm{perm}(1,\cdots,n-1)} \tr(t^{a_1}\cdots t^{a_n})
    \;C(1,\cdots,n)
  \end{align*}
In the soft limit, Berends and Giele~\cite{BeGi89} shew that the colour-stripped
amplitudes $C(1,\cdots, n)$ reduce to the corresponding amplitude without soft
gluons, times a soft factor $s_{12\dots}$:
  \begin{align*}
    C(1,\ul{2},\cdots,\ul{m},m+1,\cdots,n) &= s_{1,\ul{2},\cdots,\ul{m},m+1} C(1,m+1,\cdots,n)
  \end{align*}
We can explicitly express the soft factors $s_{i\ul{1}\ul{2}\ul{3}k}$ in terms of
the kinematical coefficients of the one gluon current and of the irreducible
correlations:
\begin{align}
  s_{i\ul{1}\ul{2}\ul{3}k} &= \gamma_i(1,2;3) + \gamma_{i}(3,2;1) 
  - \half \big[ \gamma_{i}(1,2) \,j_i(3) + j_i(1)\, \gamma_{i}(2,3) \big] \\
  &+\gamma_{i}(1,2) \,j_k(3)+ \half \,j_i(1) \,j_i(2) \,j_k(3)
  - \textstyle{\frac{1}{6}} \,j_i(1) \,j_i(2) \,j_i(3)
  - \textstyle{\binom{1 \leftrightarrow 3}{i \leftrightarrow k}}
\end{align}
where $j_k^\mu(1)\dug \frac{p_k^\mu}{p_k\cdot q_1}$.

\section{Squared soft currents\label{s:ssc}}

The objects that are relevant for cross sections involving some number of soft
gluons are the squared currents.  By using Dirac's notation~\cite{CaSe97} in
colour and helicity space, we can consider an amplitude which depends on colours
and helicities of the outgoing particles as the components of an abstract vector
in colour and helicity space. In this space, a soft current is a rectangular
matrix.  The squared amplitude is obtained by multiplying the amplitude by its
complex conjugate, and summing over final state quantum numbers.  while the
modulus square of the current is a square matrix in the space of the hard
partons' quantum numbers.  Because of the conservation of the soft gluon
currents, the square currents are explicitly gauge invariant operators when
acting on colour singlet states.

The square current for one soft gluon is just a sum of colour dipoles
$\T_i\cdot \T_k\; \dug\; \sum_a T_i^a T_k^a$ representing the colour matrices of
two hard particles summed over the same gluon colour index which connecs them:
\begin{align*}
  | \J(q) |^2 &\eqcs  -\sum_{i,k\in \mathrm{hard}} \T_i\cdot \T_k \; \S_{ik}(q)
  \ugd W(q) \;, &
  \S_{ik}(q) = \frac{p_i\cdot p_k}{p_i\cdot q\; p_k\cdot q}
\end{align*}
The square current for two soft gluons can be conveniently expressed as the
symmetric product of the square currents of a single gluon, plus an irreducible
correlation term $W(q_1,q_2)$:
\begin{align*}
  | \J(q_1,q_2) |^2 &\eqcs W(q_1)\ast W(q_2) + W(q_1,q_2) \\
  W(q_1,q_2) &=  -C_A \sum_{i,k\in\mathrm{hard}} \T_i\cdot \T_k \; {\cal S}_{ik}(q_1,q_2)
\end{align*}
This irreducible correlation is again a sum of colour dipoles, times the adjoint
Casimir, and a kinematical coefficients which has been derived
in~\cite{CaGr99}.

\begin{figure}[ht]
  {\centering
    \includegraphics[width=0.17\linewidth]{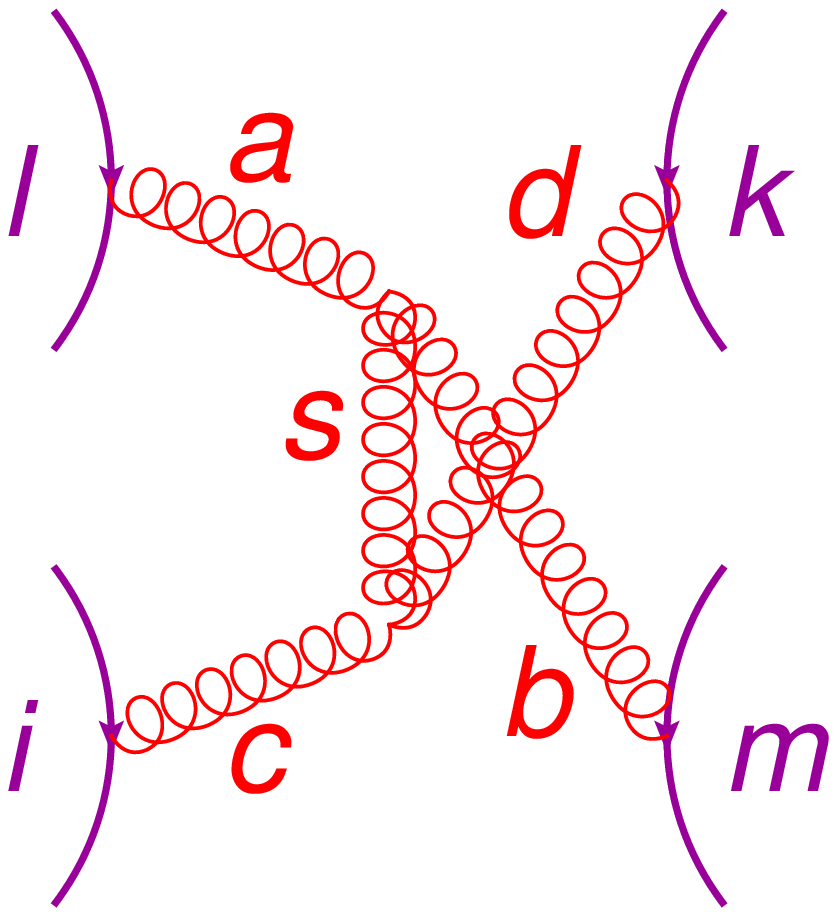}
    \hspace{0.2\linewidth}
    \includegraphics[width=0.17\linewidth]{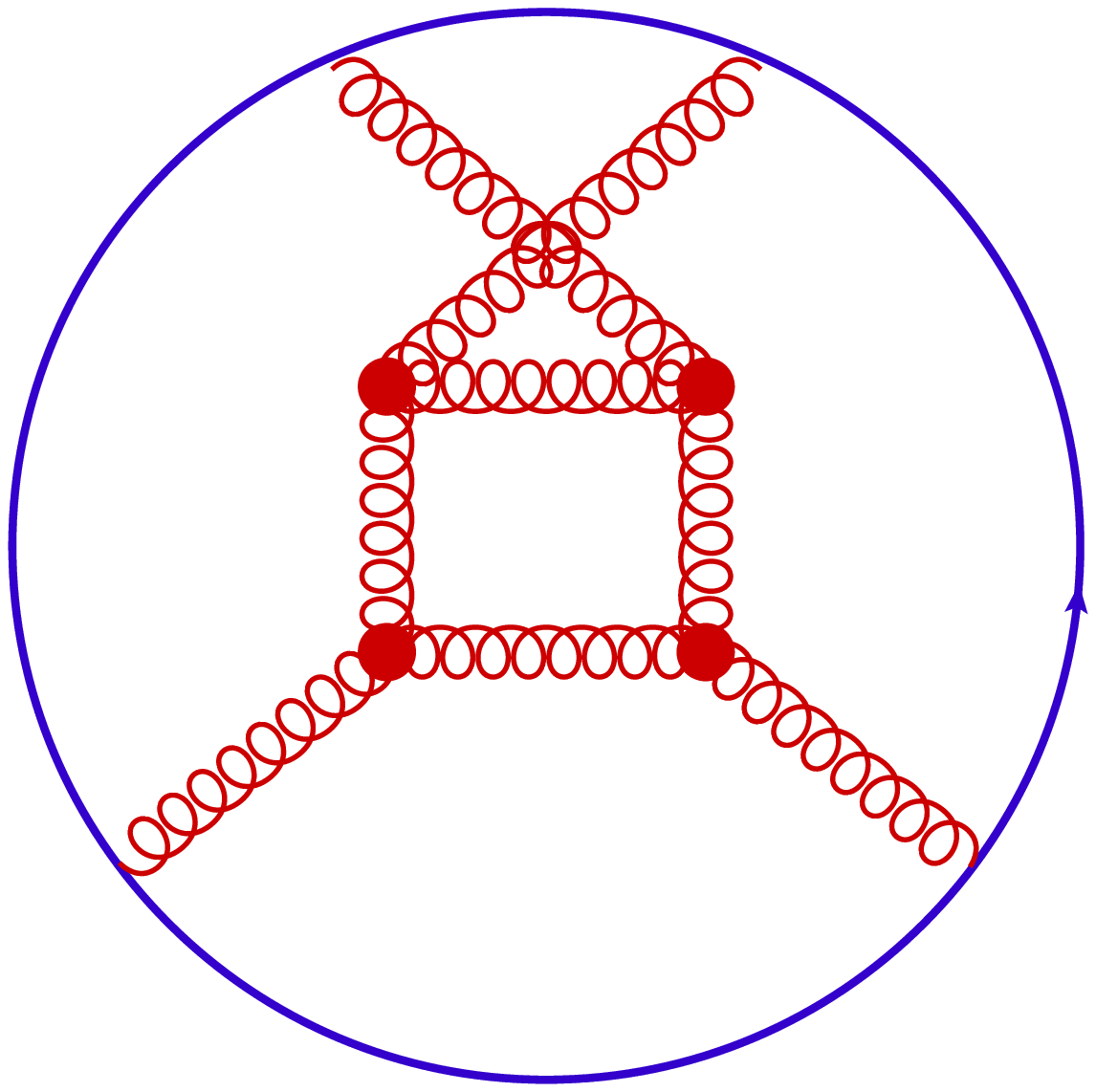}\\}
  \caption{(a-left): quadrupoles' colour structure;
    (b-right): colour monster's structure.\label{f:quadrupole}}
\end{figure}

We computed the square of the three soft gluon current. It is also conveniently
expressed as a sum of symmetric products of object with one or two soft gluons,
plus an irreducible correlation $W(1,2,3)$:
\begin{align*}
  | \J(q_1,q_2,q_3) |^2 &\eqcs W(q_1)\ast W(q_2)\ast W(q_3) \\
  &\quad + \Big[\cic W(q_1)\ast W(q_2,q_3)\Big] + W(q_1,q_2,q_3) \;.
\end{align*}
$W(q_1,q_2,q_3)$ involves not only colour dipoles but also colour quadrupoles,
the latter being four colour matrices of hard partons connected by two structure
constants, as in fig.~\ref{f:quadrupole}-a:
\begin{align}
  W(q_1,q_2,q_3) &= -C_A^2 \sum_{i,k} \T_i\cdot \T_k \;
  {\cal S}_{ik}(q_1,q_2,q_3) + \sum_{iklm} Q_{iklm} \; {\cal S}_{iklm}(q_1,q_2,q_3)\\
  Q_{iklm} &\dug \half f^{ab,cd} \Big(T_l^a\{T_i^c,T_k^d\}T_m^b\;+\;\text{h.c.}\Big)
\label{mp}
\end{align}
where we have identified a particular form of quadrupoles --- see eq.~\eqref{mp}
--- that never reduces to dipoles when some index of the hard particles are
equal.

The kinematical coefficients of the dipole and quadrupole terms are rather
cumbersome, and can be read in ref.~\cite{CCT}. They considerably simplify in
the case of strong energy ordering (SEO) of the soft gluons. In particular, the
dipole kinematical coefficient is remarkably symmetric with respect to the
permutations of the three soft momenta:
\begin{align*}
    S_{ik}^{\stro} &=
    \frac{2 (p_i \cdot p_k)^3}{3 (p_i \cdot q_1) (p_k \cdot q_1) 
      (p_i \cdot q_2) (p_k \cdot q_2) (p_i \cdot q_3) (p_k \cdot q_3)} \\
& - \frac{2 (p_i \cdot p_k)^2}{(q_1 \cdot q_2) (p_i \cdot q_1) (p_k \cdot q_2) 
(p_i \cdot q_3) (p_k \cdot q_3)} \\
& + \frac{2 p_i \cdot p_k}{(q_1 \cdot q_3) (q_2 \cdot q_3) (p_i \cdot q_1) (p_k \cdot q_2)} \;
 + {\rm perms.} \,\{1,2,3\}
\end{align*}
while the kinematical coefficient of the quadrupole is remarkably symmetric in the
exchange of the two softest momenta.

We consider now the collinear singularities of the squared currents. We all know
that the square amplitude is singular --- more precisely, not integrable ---
when momenta of two or more of external massless legs become collinear.

In the case of one soft gluon collinear to parton $B$, it is easy to show the
absence of colour correlations because of colour coherence, and the soft current
is proportional to the Casimir of the $B$ hard parton.

In the case of three soft gluons, the expansion in irreducible correlations reduces
the collinear singularities of $W$’s, so that only $W(2,3)$ and
$W(1,2,3)_\mathrm{dipole}$ are singular, according to the following scheme:
\begin{itemize}
\item $W(2,3)$ 
  \begin{itemize}
  \item[$c_1$] double-collinear limit of the two soft gluons
    (exact $P^{\mu\nu}_{g_1 g_2}$)
  \item[$c_2$] triple-collinear limit of the two soft gluons and a hard parton
  \end{itemize}
\item $W(1,2,3)_\mathrm{dipole}$
  \begin{itemize}
  \item[$c_3$] double-collinear limit of two soft gluons (exact)
  \item[$c_4$] triple-collinear limit of the three soft gluons
    (exact $P^{\mu\nu}_{g_1 g_2 g_3}$)
  \item[$c_5$] quadruple-collinear limit of the three soft gluons and a hard
    parton (soft $P^{s s'}_{g_1 g_2 g_3 C}$)
  \end{itemize}
\end{itemize}
the last result being new, to our knowledge.  $W(1,2,3)_\mathrm{quadrupole}$ has
no collinear singularity at all.

\section{Three hard partons}

Up to now, the number of hard legs was arbitrary. Now we consider the special
case of amplitudes with three hard partons $\ket{ABC}$ plus soft gluons, and
possibly other colourless particles.

Because of flavour conservation, we can have only three partonic configurations:
In the case of $\ket{gq\qb}$, there is only one colour singlet state. Therefore the
squared current, which conserves colour, acts on a one-dimensional state, so that
the soft factorization becomes a multiplication by a c-number, the eigenvalue of
the soft current on this state. This fact is valid to all perturbative orders in
the amplitudes and in the current.

In the case of three hard gluons, we can have two distinct colour-singlet
states: the colour-antisymmetic one, where the gluons are in an $\ket{f^{abc}}$
state, and the colour-symmetric $\ket{d^{abc}}$ state. Since they have opposite
charge conjugation, and the soft current conserves charge conjugation, it turns
out that the square current is a 2x2 diagonal matrix in the basis of these two
states.  The eigenvalues on these states are equal for one and two soft gluons,
but differ for three or more soft gluons, because of the quadrupole
contribution, which vanishes on the $\ket{d}$ state but not on the $\ket{f}$
state.

Without entering into many details, singlet state of three hard partons are
eigenstates of all dipole operators, and the eigenvalue is a linear combination
of Casimir coefficients.  Therefore, the one and two gluon square currents are
just c-numbers containing the Casimir of particle $A$, which is always a gluon
in our notations, and of particle $B$, which can be either a quark or a gluon.
These currents obey the so-called Casimir scaling: moving from the $\ket{gq\qb}$
state to the $\ket{ggg}$ state amounts to replace $C_F$ with $C_A$.

The same holds for the dipole terms of the three gluon square current. However,
the quadrupole term violates Casimir scaling, because of its peculiar action on
the three different hard states.  Note that, increasing the number of colours,
the dipole terms grow like $N_c^3$ while the quadrupole ones just like $N_c$,
Another peculiarity of the quadrupole term is that its kinematical coefficient
is collinear safe with respect to angular integration over soft-gluon momenta.

We have also shown that, when three hard gluons emit $N$ soft gluon with strong
energy ordering $E_1\ll E_2 \ll \dots E_N$, the squared current, to
leading order in the number of colours, has a very simple form, in terms of a
multi-eikonal function $F_{\eik}$ introduced by Bassetto-Ciafaloni-Marchesini
(BCM)~\cite{Bassetto:1984ik}:
\begin{align*}
  |\M(p_k,q_i)|^2 &\simeq |\M(p_k)|^2\; |\J(q_i)|^{2,\stro}_{\g\g\g_f}
  \left\{1+\ord{N_c^{-2}} \right\} \\
  |\J(q_i)|^{2,\stro}_{\g\g\g} &=  C_A^N \;p_A \!\cdot\! p_B\;\; p_B\!\cdot\! p_C\;\; p_C \!\cdot\! p_A \;
  F_{\eik}(p_A,p_B,p_C,q_1,\cdots,q_N) \\
  F_{\eik}(k_1,\cdots,k_M) &\dug \bigl[ 
    (k_1 \cdot k_2) (k_2\cdot k_3) \dots (k_{M-1} \cdot k_M) (k_{M} \cdot k_1)
    \bigr]^{-1}
  + \text{ineq. perm} \{k_1,\cdots,k_M\}
\end{align*}

\section{Two hard partons}

Finally, we consider soft gluon emission from two hard partons (plus any
colourless particles) in a colour-singlet state $\ket{BC}$. There are just two
such states: $\ket{q\qb}$ and $\ket{gg}$.  In both cases, the colour space is
one-dimensional and we have again c-number factorization.  Non-abelian effects
are in $SU(N_c)$ colour coefficients. The eigenvalues of the squared current on
such states obey Casimir scaling up to three soft gluons:
\begin{align*}
  |\J(q_1,q_2,q_3) |^{2}_{BC} &=
  C_B^3 \,w_{BC}(q_1) \;w_{BC}(q_2) \;w_{BC}(q_3) \\
  & \;+C_B^2 C_A [ w_{BC}(q_1) \;w_{BC}(q_2,q_3) +\text{cyc.perm.}(123)] \\
  &\;+ C_B C_A^2 \,w_{BC}(q_1,q_2,q_3)
\end{align*}
In the case of emission of $N$ soft gluons from two hard gluons, we have checked
the BCM formula in terms of the multi-eikonal function $F_{\eik}$, up to colour
suppressed contributions.  Actually, in the case of four soft gluons, we can
explicitly compute the corresponding correction, because a four soft current is
constrained by a three soft current. In fact, if one of the $N$ soft gluon is much
harder than the others, we can factorize the emission of the gluon $q_N$ from the
hard pair $BC$, and then consider the emission of the remaining $N-1$ softest gluons
from the three hard particles $q_N,B,C$. In this way, the four soft gluon
square current is constrained by the three soft gluon square current.
It turns out that the irreducible correlation for four soft gluons, in the limit
of $E_4 \gg E_{1,2,3}$, has a dipole contribution and a quadrupole
contribution. Therefore, a term satisfies Casimir scaling, while the other does
not:
\begin{align*}
  W(q_1\cdots q_4)|_{BC} =
  C_B \left[ C_A^3 \,w^{(L)}_{BC}(q_1\cdots q_4) 
    + \lambda_B N_c \,w^{(S)}_{BC}(q_1\cdots q_4)\right] 
\end{align*}
with $\lambda_F=1/2$ on a $\ket{gq\qb}$ state, $\lambda_A=3$ on an $\ket{f}$
state and $\lambda=0$ on an $\ket{d}$ state.  Actually, this colour structure
is exact, because this kinematical limit requres the computation of all Feynman
diagrams. However, we can compute the kinematical coefficients only in this
energy ordering approximation. The quadrupole term that violates Casimir scaling
is related to the quartic Casimir. Therefore, changing the two hard particles
from $q\qb$ to $gg$ can be taken into account by a generalized Casimir scaling
where, in addition to change the quadratic Casimir, we change also the quartic
one.  In the case of strong ordering, the kinematical coefficients can be
derived from the three gluon current, and we have found the first correction to
the BCM formula

If you read the book on perturbative QCD by Dokshitzer, Khoze, Mueller and
Troian~\cite{Dokshitzer:1991wu}, you may have noted that in chap.~6 they have
examined the 4-soft-gluon radiation from two massless hard partons in strong
energy ordering, and found a contribution proportional to $C_B\, N_c$ from the
so-called {\em colour monster} diagram of fig.~\ref{f:quadrupole}-b.

Our results are fully consistent with the colour monster, and are related to the
quartic Casimir
\begin{align}
  C_B \,\lambda_B \,N_c = 2 \,\frac{d_{AB}^{(4)}}{D_B} - \frac{1}{12}\, C_B\,C_A^3
\end{align}
In particular, the collinear singularities of the subleading coefficient
$w_{BC}^{(S)}$ contribute, at the inclusive level, to the soft limit
$\propto\as^4\,d_{AB}^{(4)}\,/\,\epsilon$ of the
collinear evolution kernel of the parton distribution functions
\cite{Moch:2017uml, Moch:2018wjh}.  The soft limit of the evolution kernel is
proportional to the cusp anomalous dimension that violates Casimir scaling.

\section{Conclusions}

To conclude, we have computed tree-level current for triple gluon emission in
terms of irreducible correlations for one, two and three gluons, and have
obtained explicit results for colour-ordered subamplitudes.
We have computed the tree-level squared current for three-soft-gluon emission,
and found that the 3-gluon correlation involves colour dipoles and
quadrupoles.  We checked that the collinear behaviour of the squared current is
consistent with collinear factorization and angular ordering features. In
particular, quadrupoles are collinear safe.

We have specialized our calculation to the cases of three or two hard partons:
in this cases c-number factorization holds. A remarkable results is that
quadrupoles break Casimir scaling $C_F\to C_A$ when the hard pair $q\qb$ is
replaced by $gg$. We derived a generalization of the multi-eikonal
BCM~\cite{Bassetto:1984ik} with 3 hard gluons (the original one being for just
two hard gluons).

By further specializing to the case with two hard partons, we could extended the
analysis to four-soft-gluon states. We presented the full colour structure and
the kinematical coefficients in energy ordering.  We found the
$N_c^2$-suppressed colour-monster contributon which is related to the quartic
Casimirs, and we obtained a generalization of Casimir scaling
$(C_F\to C_A\;,\quad d^{(4)}_{AF}\to d^{(4)}_{AA})$.  The Colour-monster term has
collinear singularities and contributes to the cusp anomalous dimension at
$\as^4$.
Finally, We computed the first correction $\ord{1/N_c^2}$ to the multi-eikonal
BCM formula.

\section*{Acknowledgements}

\includegraphics[width=2.2em,angle=90]{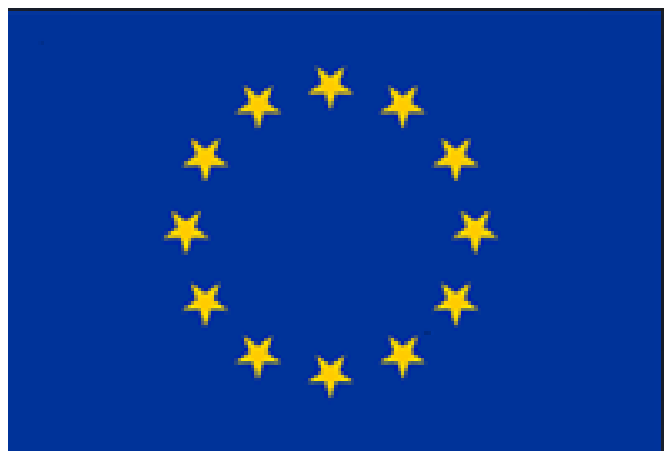}~
\begin{minipage}[b]{0.9\linewidth}
  This project has received funding from the European Union’s Horizon 2020 research and innovation programme under grant agreement No 824093.
\end{minipage}

\end{document}